\documentclass[english,twoside,a4paper,10pt]{article}

\usepackage[latin1]{inputenc}
\usepackage[T1]{fontenc}
\usepackage{amsmath}
\usepackage{amsfonts}
\usepackage{graphicx}
\usepackage{subfigure}
\usepackage{a4wide}
\usepackage{amssymb}
\usepackage{fancyhdr}
\usepackage{mathrsfs}
\usepackage[toc,page]{appendix}

\begin{document}

\title{\bf Inhomogeneous viscous fluids for inflation}
\author{ 
Ratbay Myrzakulov\footnote{Email: rmyrzakulov@gmail.com},\,\,\,
Lorenzo Sebastiani\footnote{E-mail address: l.sebastiani@science.unitn.it
}\\
\\
\begin{small}
Eurasian International Center for Theoretical Physics and  Department of General
\end{small}\\
\begin{small} 
Theoretical Physics, Eurasian National University, Astana 010008, Kazakhstan
\end{small}\\
}

\date{}

\maketitle


\begin{abstract}
In this paper, we investigate inhomogeneous viscous fluid cosmology for inflation. Several toy models are presented in the attempt to analyze how inflation can be realized according with cosmological data by making use of an inhomogeneous EoS parameter for the fluid and/or by introducing a viscosity to have a graceful exit from inflation. The results will be compared with the ones of scalar field representation and discussed. We will pay attention on the possibility to recover the reheating and therefore the Friedmann universe.
\end{abstract}



\tableofcontents
\section{Introduction}

The discovery of the accelerated expansion of the universe~\cite{WMAP}, and other evidences related to the existence of an early-time acceleration after the Big Bang, namely the inflation~\cite{Linde, revinflazione}, suggest the presence of `dark' fluids different to standard matter and radiation at the cosmological level. 
The origin of these dark universe contents remains unknown, and they may come from the theory of gravitation itself (modified theories of gravity~\cite{reviewmod, reviewmod2, myrev}), be in the form of the Cosmological constant (the simplest explanation for the current acceleration), have a scalar fields representation (the inflaton in the inflationary scenario), or be the effect of some non-perfect fluids.

In this paper, we focused our attention on the early-time acceleration that universe underwent at the time of the Big Bang. Inflation has been stated to solve the problems of the initial conditions of Friedmann universe (horizon problem, flatness problem): moreover, it could explain some issue related to the particle physics (monopole magnetic problem). Despite to the fact that the observations of the inhomogeneities in the present universe furnish several informations  about the viable scenario of the primordial acceleration, the choice of the models is quite large.

The most popular inflation models are based on scalar fields representation: an homogeneous field (the inflaton), produces the acceleration under some suitable conditions. Following the first proposal of Guth~\cite{Guth} and Sato~\cite{Sato}, in the last years many of this kind of models have been investigated.
Typically, the magnitude of the inflaton is very large at the beginning of the inflation,
with several possibilities for the initial boundary conditions (``chaotic'' inflation~\cite{chaotic}). 
At the end of the inflation the 
inflaton falls in a potential hole and starts to oscillate, such that the 
rehating processes take place~\cite{buca1, buca2, buca3, buca4}.
Some more complicated models are based on a phase transition between two scalar 
fields (hybrid or double inflation models~\cite{ibrida1, 
ibrida2}).

We also mention that until today, cosmological and astrophysical data seem to confirm 
the predictions of the so called Starobinsky model~\cite{Staro},
based on the account of $R^2$-term as a correction of the Einstein equations: this model has a corresponding viable inflation in the scalar field framework according with the Planck data~\cite{Planckdata}, even if the very recent experiments on microwave radiation~\cite{BICEP2} seem to indicate some discrepances (see also Refs.~\cite{Staromio, Staromio2}). 

In this context, we would like to investigate some features of inhomogeneous viscous fluids in hot universe scenario, namely how inflation can be reproduced by using an inhomogeneous Equation of State parameter and/or some viscosity for non-perfect fluids driving primordial acceleration. The end of inflation will be also analyzed, with some consideration on the rehating phase and the connection between exit from inflation and Friedmann universe. 

The paper is organized as follows. In Section {\bf 2}, we give a brief review of inflation in scalar field theories. In Section {\bf 3}, the formalism of inhomogeneous viscous fluids in flat Friedmann-Robertson- Walker space-time is presented and inflation in several toy models is investigated. In our analysis, we will try to give an exhaustive description of inflation 
induced by non-constant EoS parameter or by some viscosity which renders possible a graceful exit in Friedmann universe. We will see how the description of a viable inflation changes in inhomogeneous viscous fluid representation with respect to the scalar field one, and we will pay attention on the end of inflation: in the specific, the rehating phase as the production of matter particles or the conversion of the fluid energy in standard radiation are studied. In Section {\bf 4}, to complete the work, we will see how it could be possible to unify in a unique fluid model the early-time with the late-time acceleration. Conclusions are given in Section {\bf 5}.

We use units of $k_{\mathrm{B}} = c = \hbar = 1$ and denote the
gravitational constant, $G_N$, by $\kappa^2\equiv 8 \pi G_{N}$, such that
$G_{N}^{-1/2} =M_{\mathrm{Pl}}$, $M_{\mathrm{Pl}} =1.2 \times 10^{19}$ GeV being the Planck mass.


\section{Inflation in scalar field theories}

Let us review the dynamics of inflation in scalar field theories. The action of a canonical scalar field theory in curved space-time is given by
\begin{equation}
I=\int_{\mathcal{M}} d^4 x \sqrt{-g} \left( 
\frac{R}{2\kappa^2} -
\frac{1}{2}g^{\mu\nu}
\partial_\mu \sigma \partial_\nu \sigma - 
V(\sigma)\right)\,,
\end{equation}
where $\sigma$ is the scalar field subjected to the potential $V(\sigma)$, $g$ is the determinant of the metric tensor $g_{\mu\nu}$, $R$ is the Ricci scalar and $\mathcal M$ is the space-time manifold. We will use the flat Friedmann-Robertson-Walker (FRW) metric,
\begin{equation}
ds^2=-dt^2+a(t)^2d{\bf x}^2\,,\label{metric}
\end{equation}
where $a(t)$ is the scale factor of the universe. The (homogeneous) scalar field $\sigma$, which is  identified with the inflaton, is  a function of the cosmological time only. Thus,
the equations of motion (EOMs) are derived from the action as
\begin{equation}
\frac{3 
H^2}{\kappa^2}=\rho_\sigma\,,\quad-\frac{1}{\kappa^2}\left(2\dot 
H+3 H^2\right)=p_\sigma\,,
\end{equation}
where $H(t)$=$\dot a(t)/a(t)$ is the Hubble parameter, the dot being the derivative with respect to the time, and $\rho_\sigma$ and $p_\sigma$ are the energy density and the pressure of the field $\sigma$, respectively
\begin{equation}
\rho_\sigma=\frac{\dot\sigma^2}{2}+V(\sigma)\,,\quad 
p_\sigma=\frac{\dot\sigma^2}{2}-V(\sigma)\,.
\end{equation}
By combining the EOMs we also obtain the conservation law,
\begin{equation}
\dot\rho_\sigma+3H(\rho_\sigma+p_\sigma)=0\,.
\end{equation}
Explicitly, the equations of motion read
\begin{equation}
\frac{3H^2}{\kappa^2}=\frac{\dot\sigma^2}{2}+V(\sigma)\,,\quad 
-\frac{1}{\kappa^2}\left(2\dot 
H+3 H^2\right)=\frac{\dot\sigma^2}{2}-V(\sigma)\,,
\end{equation}
and the energy conservation law coincides with the equation of motion for 
$\sigma$,
\begin{equation}
\ddot\sigma+3H\dot\sigma=-V'(\sigma)\,,
\end{equation}
where the prime denotes the derivative of the potential with respect to 
the field.

The acceleration can be evaluated as
\begin{equation}
\frac{\ddot{a}}{a}= H^2+\dot{H}=H^2\left(1-\epsilon \right)\,,
\end{equation}
where we have introduced the so called ``slow roll'' parameter
\begin{equation}
\epsilon=-\frac{\dot{H}}{H^2}\,.\label{epsilon}
\end{equation}
Thus, in order to have an acceleration, one must require $\epsilon<1$. An other important slow roll parameter in studying inflation is given by
\begin{equation}
\eta=-\frac{\ddot{H}}{2H\dot{H}}=\epsilon-\frac{1}{2 \epsilon H}\dot\epsilon\,.\label{eta}
\end{equation}
Generally speaking, the inflation is described by a de Sitter expansion, but, due to the perturbation of the field, the Hubble parameter slowly decreases and finally the inflation ends. The mechanism is the following. At the beginning, the field, which in the chaotic inflation is negative and very large, is in the ``slow roll''  regime,
\begin{equation}
\dot\sigma^2\ll V(\sigma)\,,
\end{equation}
namely its kinetic energy has to be small with respect to the potential. As 
a consequence, the field EoS parameter results to be
\begin{equation}
\omega_\sigma\equiv\frac{p_\sigma}{\rho_\sigma}=\frac{\dot\sigma-2V(\sigma)}{\dot\sigma+2V(\sigma)}\simeq-1\,,
\end{equation}
and the expansion is governed by the de Sitter solution with Hubble parameter  
\begin{equation}
\frac{3H^2}{\kappa^2}\simeq V(\sigma)\,.
\end{equation}
On the other hand, the magnitude of the (negative) field must slowly increases as
\begin{equation}
3H\dot\sigma\simeq -V'(\sigma)\,,
\end{equation}
assuming $V'(\sigma)>0$. Therefore, the inflaton rolls down towards a potential minimum where the inflation ends. 
It is easy to understand that in the slow roll regime $\epsilon\ll 1$ and $|\eta|\ll1$, and inflation ends when this paramters become of the order of the unit (when $\epsilon=1$, the acceleration goes to zero). After that, the inflaton starts to oscillate and the rehating processes take place.\\
\\
The primordial acceleration can solve the problems of initial conditions of the universe (horizon and velocities problems), only if 
\begin{equation}
\dot a_\mathrm{i}/\dot a_0< 10^{-5}\,, \label{N0}
\end{equation}
where $\dot a_\mathrm{i}\,,\dot a_0$ are the time derivatives of the scale factor at the Big Bang and today, respectively, and $10^{-5}$ is the estimated value of the inhomogeneous cosmological perturbations. Since during radiation/matter era $\dot a(t)$ only decreases of a factor $10^{28}$, it is required that $\dot a_\mathrm{i}/\dot a_\mathrm{f}<10^{-33}$, where $a_\mathrm{i}$ is the scale factor at the Big Bang (it means, at the beginning of inflation), and $a_\mathrm{f}$ is the scale factor at the end of inflation. Furthermore, for a de Sitter expansion where $\dot a_\mathrm{f}/\dot a_\mathrm{i}\equiv a_\mathrm{f}/ a_\mathrm{i}$, we may introduce the number of $e$-folds $N$ as
\begin{equation}
N\equiv\ln \left(\frac{a_\mathrm{f}}{a_\mathrm{i}}\right)
=\int^{t_f}_{t_i} H(t)dt\,,\label{N}
\end{equation}
and inflation is viable when $N>76$. 

The amplitude of the primordial scalar power spectrum is given by
\begin{equation}
\Delta_{\mathcal R}^2=\frac{\kappa^2 H^2}{8\pi^2\epsilon}\,,\label{spectrum}
\end{equation}
and from slow roll paramters one gets the spectral index $n_s$ and the tensor-to-scalar 
ratio $r$,
\begin{equation}
n_s=1-6\epsilon+2\eta\,,\quad r=16\epsilon\,.\label{index}
\end{equation}
The last results observed by the Planck satellite are 
$n_{\mathrm{s}} = 0.9603 \pm 0.0073\, (68\%\,\mathrm{CL})$ and 
$r < 0.11\, (95\%\,\mathrm{CL})$. In general, the inflation models built
on flat potentials, well satisfy this bounds. However,
very recently, 
the BICEP2 experiment~\cite{BICEP2} has detected the $B$-mode polarization of the cosmic microwave background (CMB) radiation with the tensor to scalar ratio 
\begin{equation}
r =0.20_{-0.05}^{+0.07}\, (68\%\,\mathrm{CL})\,,\label{r}
\end{equation}  
and the case that $r$ vanishes 
has been rejected at $7.0 \sigma$ level. 
Thus, the viable models of inflation may have to produce such a finite value of 
$r$.

\section{Fluid models for inflation and reheating}

Let us consider an inhomogeneus viscous fluid whose Equation of State (EoS) assumes the general form~\cite{fluidOd, fluidOd2, fluidOd3, fluidOd4},
\begin{equation}
p=\omega(\rho)\rho-3 H\zeta(a,H,\dot H, \ddot H,...)\,.\label{EoS}
\end{equation}
Here, $\zeta(a, H,...)$ is the bulk viscosity, and may depend on the scale factor, the Hubble parameter and the derivatives of Hubble parameter. Also the EoS parameter of fluid, $\omega(\rho)$, may be not a constant and has a dependence on the energy density in the case of inhomogeneous fluid.
Moreover, in order to obtain the positive sign of the entropy change in an irreversible process, $\zeta(a,H,..)$ has to be positive~\cite{Alessia, Brevik,Alessia(2)}. 

By introducing this fluid in the background of General Realtivity, the Friedmann equations for flat FRW metric (\ref{metric}) read
\begin{equation}
\rho=\frac{3}{\kappa^{2}}H^{2}\,,
\quad
p=-\frac{1}{\kappa^{2}} \left( 2\dot H+3H^{2} \right)\,,\label{F2}
\end{equation}
and the energy conservation law of the fluid assumes the form
\begin{equation}
\dot\rho+3H\rho(1+\omega(\rho))=(3H)^2\zeta(a, H, \dot H, \ddot H,...)\,.\label{cons2}
\end{equation}
We want to stress that the formalism of viscous fluid cosmology may be extended to other theories for inflation. In particular, almost any modified gravity theory can be encoded in a fluid-like form to get at least some intermediate useful results~\cite{Caprev}. For example, for $F(R)$-gravity whose lagrangian is given by $\mathcal L=F(R)$, $F(R)$ being a function of the Ricci scalar only, 
the FRW equations of motion can be written in the usual Friedmann-like form (\ref{F2})
by introducing an effective viscous fluid whose effective energy density and pressure read
\begin{eqnarray}
\rho_{\mathrm{eff}} &\equiv&
\frac{1}{2\kappa^{2}}
\left[ \left( F_R(R) R-F(R) \right)-6H^2(F_R(R)-1)
-6H\dot{F}_R(R)
\right]\,
\label{rhoeffRG} \\ \nonumber\\
p_{\mathrm{eff}} &\equiv&
\frac{1}{2\kappa^{2}} \Bigl[
-\left( F_R(R)R-F(R)\right)+(4\dot{H}+6H^2)(F_R(R)-1)
+4H \dot F_R(R)+2\ddot{F}_R(R)
\Bigr]\,, \nonumber\\
\end{eqnarray}
and obey to the EoS (\ref{EoS}) with $\omega=-1$ and
\begin{eqnarray}
\zeta(H,\dot H, \ddot H,...)&\equiv&-\frac{1}{6H\kappa^2}\left(4\dot H(F_R(R)-1)-2H \dot F_R(R)+2\ddot F_R(R)\right)\,.
\end{eqnarray} 
In the above expressions, the pedex `$R$' indicates the derivative with respect to $R$.

Let us see some instructive examples of inhomogeneus viscous fluid cosmology applied to inflation.

\subsection{Fluid model with $\omega(\rho)=-\frac{\rho}{\rho+\rho_*}$ for inflation and reheating\label{secomega}}

Let us start from an inhomogeneous fluid with zero viscosity ($\zeta(a, H,...)=0$) and EoS parameter in the form
\begin{equation}
\omega(\rho)=-\frac{\rho}{\rho+\rho_*}\,,\label{o1}
\end{equation}
where $\rho_*$ is an (effective) energy density at the scale of inflation ($\rho_*\sim H_*^2/\kappa^2$). We get
\begin{equation}
\omega(1\ll\rho/\rho_*)\simeq-1\,,\quad \omega(\rho/\rho_*\ll 1)\simeq0\,.
\end{equation}
When $1\ll\rho/\rho_*$ (it means, $\kappa^2\rho_*\ll H^2$), we get the de Sitter accelerated expansion of inflation,
\begin{equation}
H=H_0\,,
\end{equation}
where $H_0$ is a constant fixed from the boundary conditions of the fluid and in general $H_0\ll M_{Pl}$ to avoid quantum corrections to gravity, $M_{Pl}$ being the Planck mass.

From the conservation law (\ref{cons2}) we also obtain
\begin{equation}
\frac{\dot\rho}{\rho}\simeq -3H\left(\frac{\rho_*}{\rho+\rho_*}\right)\simeq -3H\frac{\rho_*}{\rho}\,,\label{consinfl}
\end{equation}
such that we finally have
\begin{equation}
\rho=\frac{3H_0^2}{\kappa^2}-3\rho_*\log\left[a(t)\right]\simeq
\frac{3H_0^2}{\kappa^2}+3H_0\rho_*(t_i-t)+\frac{3\rho_*^2\kappa^2(t_i-t)^2}{4}\,,
\label{rr}
\end{equation}
where we have set $\rho=3H_0^2/\kappa^2$ at $t=t_i$, $t_i$ being the initial time of inflation, and we also have implemented the following expression of the scale factor,
\begin{equation}
a(t)=\exp\left[H_0(t-t_i)-\frac{\kappa^2\rho_*}{4}(t_i-t)^2\right]\,,
\end{equation}
which has been derived by taking $\rho$ at the first order of $(t_0-t)$ and therefore by integrating the first equation in (\ref{F2}).
We normalized $a(t_i)=1$, since any constant factor in front of $a(t)$ can be absorbed by the constant term of $\rho$ above due to a rescaling of it.
The universe is in a de Sitter phase as soon as
$(t-t_i)\kappa^2\rho_*\ll H_0$, but when the time increases, the energy density of fluid decreases and the universe exits from inflation. Thus, the duration of inflation can be estimated as $\Delta t\simeq H_0/(\kappa^2\rho_*)$, and since the $N$-folds number (\ref{N}) reads $N\simeq H_0\Delta t$, one finally has
\begin{equation}
\kappa^2\rho_*\simeq H_0^2/N\,.
\end{equation}
The de Sitter Hubble paramter $H_0$ depends on the initial value of the fluid $\rho_i(=3H_0/\kappa^2)$, and at the beginning of inflation must be
\begin{equation}
\rho_i\simeq3N\rho_*\,. \label{rhoN}
\end{equation}
By taking into account that $N>76$, we get $\rho_i>228\rho_*$. This result can be directly aquired by using (\ref{N}) which leads to
\begin{equation}
N\equiv\int^{t_f}_{t_i} H(t)dt=\int^{\rho_f}_{\rho_i}\frac{H}{\dot\rho}d\rho
=\frac{1}{3\rho_*}(\rho_i-\rho_f)\simeq\frac{\rho_i}{3\rho_*}\,,
\end{equation}
in agreement with (\ref{rhoN}).
The slow roll parameters (\ref{epsilon})--(\ref{eta}) are given by
\begin{equation}
\epsilon\simeq\frac{3\rho_*}{2\rho}\simeq\frac{1}{2N} 
\,,\quad
\eta\simeq\frac{3\rho_*}{4\rho}\simeq\frac{1}{4N}\,,
\end{equation}
where in order to evaluate $\eta$ we have used (\ref{rr}) in the first equation of (\ref{F2}).
Some remarks are in order. For a fluid model, since $N\simeq H_i\Delta t$ and $\Delta t\simeq t_f\equiv\Delta\rho/\dot\rho\simeq\rho_i/\dot\rho_i$, by considering that $\rho/\dot\rho\sim H/\dot H$, the $N$-folds number results to be $N\sim H_i^2/\dot H_i\simeq 1/\epsilon$ and we have a viable inflation if $t_f>76 H_i^{-1}$. In the scalar tensor theories, $\Delta t\equiv\Delta\sigma/\dot\sigma\simeq\sigma_i/\dot\sigma_i$, but $\sigma/\dot\sigma\sim\sqrt{H^2/\dot H}(\sigma/H)$ and $N\sim\sqrt{H_i^2/\dot H_i}\sigma_i\sim1/\sqrt{\epsilon}$, which leads to a smaller value of $\epsilon$ when $N$ is given ($\epsilon\sim1/N^2$).

The power spectrum (\ref{spectrum}) and the spectral indexes (\ref{index}) read
\begin{equation}
\Delta_\mathcal R=\frac{\kappa^4\rho^2}{36\pi^2\rho_*}\,,\quad
n_s=1-\frac{15\rho_*}{2\rho}\simeq1-\frac{15}{6N}\,,
\quad
r=\frac{24\rho_*}{\rho}\simeq\frac{8}{N}\,.
\end{equation}
For example, for $N=76$ (which corresponds to an initial value of the field  $\rho_i=228\rho_*$), we obtain $n_s=\simeq0.967$ and $r\simeq0.105$. The fluid model permits to have a tensor-to-scalar ratio $r$ much largr with respect to the scalar field models where $r\sim N^{-2}$, such that we still are in agreement with Plank data and closer to the BICEP2 results (\ref{r}).\\
\\
Let us see now what happens at the end of inflation, in the limit $\rho\ll\rho_*$. In this case, $\omega(\rho)\simeq 0$, but in principle the model can be easly rewritten for any desired EoS fluid-parameter after inflation. If we want to obtain a generic $\omega=\omega_\text{eff}$, $\omega_\text{eff}$ being a constant, we redefine (\ref{o1}) as
\begin{equation}
\omega(\rho)=-(\omega_\text{eff}+1)\frac{\rho}{\rho+\rho_*}+\omega_\text{eff}\,,\quad
\omega_\text{eff}\neq -1\,,\label{ooooo}
\end{equation}
such that $\omega\simeq-1$ when $\rho_*\ll\rho$ and $\omega\simeq\omega_\text{eff}$ when $\rho\ll\rho_*$. The analysis of inflation is the same of above, since Equation (\ref{consinfl}) is still valid. However, a perfect fluid in expanding universe cannot lead to the rehating process after inflation: therefore, the reproduction of the standard radiation and matter dominated universe is not possible, being all the contents of the universe shifted away during the strong accelerated expansion of inflation.

In the scalar field theories, after the inflation the inflaton starts to oscillate. Thus, due to a coupling between the inflaton and the matter field, the creation of particles takes place during this oscillations. Here, the coupling between inflaton and matter field must be replaced by a coupling between curvature and matter field and, as it has been shown in Ref.~\cite{deFelice}, if the Ricci scalar oscillates with decreasing amplitude, the rehating is possible.

To obtain the oscillations of the Ricci scalar, we must modified the energy conservation law of our fluid. For the sake of simplicity,
we take the case $\omega\simeq\omega_\text{eff}=0$, namely (\ref{o1}) when $\rho\ll\rho_*$, and we introduce a viscosity term as in (\ref{EoS}) in the following way,
\begin{equation}
p=-\frac{\rho}{\rho+\rho_*}\rho-3 H\zeta(H,\dot H, \ddot H)\,,
\end{equation}
where
\begin{equation}
\zeta(H, \dot H, \ddot H)=\text{e}^{-H^2/H_*^2}f(H,\dot H, \ddot H)\,.
\end{equation}
Here, $H_*^2$ corresponds to $H_*^2=\kappa^2\rho_*/3$, such that the viscosity vanishes during inflation, when $\rho_*\ll\rho$, and tends to $-3Hf(H,\dot H, \ddot H)$ when $\rho\ll\rho_*$. 
For our purpose, we assume the following form of $f(H,\dot H, \ddot H)$,
\begin{equation}
f(H,\dot H, \ddot H)=-\frac{4}{3H^2\kappa^2\gamma}\left(\dot H^2+ H\ddot H\right)
+\frac{H}{\kappa^2}-\frac{(16 m^2+\gamma^2)}{4\gamma\kappa^2}+\frac{2\dot H^2}{\gamma\kappa^2 H^2}\,,\quad \gamma>0\,,\label{ff}
\end{equation}
where $\gamma$ and $m$ are dimensional ($[\gamma]=[m]=[H]$) positive constants. As a consequence,
when $\rho\ll\rho_*$, equation (\ref{cons2}) reads
\begin{equation}
\dot\rho+3H\rho=-\frac{12}{\kappa^2\gamma}\left(\dot H^2+ H\ddot H\right)+\frac{9}{\kappa^2}H^3-\left(\frac{16m^2+\gamma^2}{2\gamma}\right)\left(\frac{3H^2}{\kappa^2}\right)+\frac{18\dot H^2}{\gamma\kappa^2}\,.
\end{equation}
Thus, by using the first equation in (\ref{F2}), we get
\begin{equation}
\ddot\rho+\frac{\gamma}{2}\dot\rho+\left[\frac{16m^2+\gamma^2}{4}\right]\rho=\frac{3\dot\rho^2}{4\rho}\,,
\end{equation}
whose solution is given by
\begin{equation}
\rho=\rho_0\text{e}^{-\gamma\,t}\cos^4 mt\,,
\end{equation}
where $\rho_0$ is a positive constant. It follows from this expression,
\begin{equation}
H=\sqrt{\frac{\kappa^2\rho_0}{3}}\text{e}^{-\frac{\gamma\,t}{2}}\cos^2 mt\,,
\end{equation}
\begin{eqnarray}
R\equiv12H^2+6\dot H &=&
4\kappa^2\rho_0\text{e}^{-\gamma t}\cos^4 mt
-6\sqrt{\frac{\kappa^2\rho_0}{3}}\text{e}^{-\frac{\gamma t}{2}}\left[\frac{\gamma}{2}\cos^2 mt+2 m\cos mt\sin mt\right]
\nonumber\\&\simeq&
-6\sqrt{\frac{\kappa^2\rho_0}{3}}\text{e}^{-\frac{\gamma t}{2}}\left[\frac{\gamma}{2}\cos^2 mt+2 m\cos mt\sin mt\right]\,,
\end{eqnarray}
where we have considered $1\ll\gamma t$, namely $H^2\ll |\dot H|$. The scale factor behaves as
\begin{equation}
a(t)\equiv a_0\exp\left[\int H(t')dt'\right]=a_0\exp\left[-\sqrt{\frac{\kappa^2\rho_0}{3}}\frac{\text{e}^{\frac{-\gamma t}{2}}(16m^2+2\gamma^2\cos^2 mt -4m\gamma\sin 2mt}{16m^2\gamma+\gamma^3}\right]\,,
\end{equation}
where $a_0>0$ is a generic constant. The solution is for expanding universe ($H>0$, $a(t)$ real and positive) and the energy density of the fluid decreases with the expansion as
$\rho\sim\log [a(t)/a_0]^2>0$. The reheating mechanism at the origin of the particle production takes place during the fluid-dominated stage. In this case, we can write the following equation for a (bosonic) particle field $\chi$ with mass $m_\chi$,
\begin{equation}
\Box\chi-m^2_\chi\chi-\xi R\chi=0\,,
\end{equation}
where we introduced a coupling between the particle field and the Ricci scalar through the coupling constant $\xi$. For a single mode of $\chi$ with momentum $k$, namely $\chi_k$, this equation becomes on FRW space-time
\begin{equation}
\ddot\chi_k+3H\dot\chi_k+\left(m^2_\chi+\xi R\right)\chi_k=0\,.
\end{equation}
Following Ref.~\cite{deFelice}, we introduce the conformal time $\eta=\int a(t)^{-1}dt$ and the field $u_k\equiv a(t)\chi_k$, and we derive
\begin{equation}
\frac{d^2}{d\eta^2}u_k+m_\text{eff}^2 a(t)^2u_k=0\,,\label{u}
\end{equation}
where $m_\text{eff}$ is an effective mass,
\begin{equation}
m^2_\text{eff}=\left[m_\chi^2+\left(\xi-\frac{1}{6}\right)R\right]\,.\label{meff}
\end{equation}
In this way, since the effective mass of $u_\kappa$ depends on the Ricci scalar, the solution of equation (\ref{u}), and therefore $\chi_k$, namely the number of massive particles, changes with the time. In this case  it is possible to obtain the rehating during the oscillations of the Ricci scalar~\cite{deFelice}, as soon as the solution is supported (in our case) by the viscous fluid. After the particles production, it is expected that the universe enters in the radiation era and the Friedmann universe is reproduced. As a final remark, it is interesting to note that also in the case
$\xi=0$ the effective mass (\ref{meff}) still depends on $R$.

\subsection{Fluid model with viscosity $\zeta(H)=\text{e}^{-(H/H_*)}f(H)$ for inflation and radiation era\label{secviscous}} 

In this Subsection, we would like to see 
how the viscosity may realize a graceful exit from inflation.
Differently from the previous case, we assume $\omega(\rho)\equiv\omega\neq-1$ constant in (\ref{EoS}).
Inflation is realized when the viscosity is negligible, and we take for viscosity the following form
\begin{equation}
\zeta(H)=\text{e}^{-(H/H_*)}f(H)\,,\label{toyzeta}
\end{equation}
where $H_*$ is a constant Hubble parameter at the end of inflation and $f(H)$ is a suitable function to be determined. One has
\begin{equation}
\zeta(1\ll(H/H_*))\simeq 0\,,\quad\zeta((H/H_*)\ll 1)\simeq f(H)\,.
\end{equation}
We must analyze the model in the two asymptotic limits. 

When $1\ll (H/H_*)$, the solution of the Friedmann equations (\ref{F2}) is given by
\begin{equation}
H(t)=\frac{2}{3(1+\omega)t}\,,\quad\rho(t)=\rho_0a(t)^{-3(1+\omega)}\,,\quad-1<\omega\,,
\label{a}
\end{equation}
$a_0\,,\rho_0$ being constants eventually related to each other. We exclude the case $\omega<-1$, for which, in order to maintain the positivity of the Hubble parameter, we shall introduce an integration constant as
\begin{equation}
H(t)=-\frac{2}{3(1+\omega)(t_0-t)}\,,\quad\omega<-1\,,\label{b}
\end{equation}
where $0<t_0$ is a fixed time and $t<t_0$. However, in such a case, $0<\dot H$ and the Hubble parameter increases, making impossible the exit from inflation with the viscosity.

The solution (\ref{a}) shows an initial singularity at $t=0$, which can be identify with the Big Bang. The acceleration is realized in the quintessence region,
\begin{equation}
-1<\omega<-1/3\,,\quad 0<\frac{\ddot a}{a}\equiv H^2+\dot H=
\frac{4-6(1+\omega)}{9(1+\omega)^2t^2}\,.
\end{equation}
In order to correctly reproduce inflation, the solution must be close to the de Sitter one (i.e. $\omega$ close to minus one).

In the limit $(H/H_*)\ll1$, the viscous term grows up and by combining the first Friedmann equation in (\ref{F2}) with the continuity equation (\ref{cons2}), we get
\begin{equation}
\dot\rho+3H\rho\left(1+\omega-\frac{\kappa^2}{3H}f(H)\right)=0\,.
\end{equation}
Thus, if $f(H)$ is constructed in the following way,
\begin{equation}
f(H)=\frac{3H}{\kappa^2}(\omega-\omega_{\text{eff}})\,,
\end{equation}
at the end of inflation the fluid turns out to be a (perfect) fluid with EoS paramter 
$\omega_\text{eff}$:
by putting $\omega_\text{eff}=1/3$, we may recover the radiation/ultrarelativistic matter universe of the Standard Model without invoking the reheating, since is the energy density of fluid itself wich converts in radiation. 

The slow roll paramters $\epsilon$ (\ref{epsilon}) and $\eta$ (\ref{eta}) for solution (\ref{a}) are given by
\begin{equation}
\epsilon=\eta=\frac{3(1+\omega)}{2}\,,\label{ss}
\end{equation}
and we see that, as we stated above, inflation is viable only if $\omega$ is close to minus one and $\epsilon, \eta\ll1$. To measure the $N$-folds number we cannot use such paramters, since they are constant. It does not mean that the quintessence solution is stable, since the viscous term (\ref{toyzeta}) slowly changes with the decreasing of the Hubble parameter, making at some point the expressions (\ref{ss}) not still valid.
We could reasonably assume that the universe exits from inflation when $H/H_*\sim 1$. Since we are not dealing with a de Sitter solution, the $N$-folds number must be derived from (\ref{N0}) as
\begin{equation}
N\equiv\log\left(\frac{\dot a_f}{\dot a_i}\right)=\frac{2-3(1+\omega)}{3(1+\omega)}\ln\left(\frac{t_*}{t_i}\right)\,,\label{last}
\end{equation}
where $t_i$, as usually, is the time at the beginning of inflation and $t_*$ is choosen like $H_*=2/[3(1+\omega)t_*]$. This expression corresponds to the one in (\ref{N}) when $\omega$ is close to minus one.
We note that in this description the $N$-folds number, and therefore the time of inflation, do not depend on the parameters which describe the quasi de Sitter expansion. In this way, we can reconstruct every values for spectral indexes. The power spectrum (\ref{spectrum}) results to be
\begin{equation}
\Delta_\mathcal{R}^2=\frac{\kappa^2}{27\pi^2(1+\omega)^3t^2}\,,
\end{equation} 
and for spectral indexes (\ref{index}) one gets
\begin{equation}
n_s=(1-6(1+\omega))\,,\quad r=24(1+\omega)\,.
\end{equation}
We see that in order to satisfy the last BICEP2 results (\ref{r}), we could require $\omega\simeq-0.992>-1$. In this case, the $\epsilon, \eta\simeq 0.012$ slow roll parameters
(\ref{ss}) remain very small, being the solution a quasi De Sitter. Note that in the above expressions, $(1+\omega)>0$.

\subsection{Fluid model with $\omega=-1+a_1\rho^{1/2}-a_2\rho^{-1/2}$ for inflation\label{seclast}}

As a last example, we would like to present a model of fluid whose inhomogeneous EoS parameter brings the universe to expand with a quasi de Sitter solution (look also Ref.~\cite{davood}) reproducing the phenomenology of inflation. In the previous case, a viscosity term was necessary to change the Equation of State of the fluid and lead to the exit from inflation. 
Here, the effective EoS parameter of the fluid itself slowly changes during inflation and brings to a decelerated phase. Moreover, the presented solution will be an exact solution of the model, without making use of any approximation.

Let us assume the following (suitable) class of Equations of State (\ref{EoS}) for inhomogeneous fluid,
\begin{equation}
p=\left(\frac{A_0}{\sqrt{3\kappa^2}}\rho^{n-3/2}-\frac{B_0}{\sqrt{3\kappa^2\rho}}-1\right)\rho\,,\label{ppp}
\end{equation}
where $A_0\,,B_0$ are dimensional constants ($[B_0]=[H]$) and $0<n$ is a positive number. 
By combining the conservation law (\ref{cons2}) with the first Friedmann equation in (\ref{F2}), we get 
\begin{equation}
\dot\rho+A_0\rho^n=B_0\rho\,,\quad n>0\,.
\end{equation}
The solution of this equation may describe a quasi-de Sitter expansion for inflation. As an example, let us take $n=2$ (such that $[A_0]=[\kappa^2/H]$). We obtain for expanding universe ($H>0$),
\begin{equation}
\rho=\frac{B_0}{C_0\text{e}^{B_0(t_i-t)}+A_0}\,,
\quad
H=\sqrt{\frac{\kappa^2}{3}}\sqrt{\frac{B_0}{C_0\text{e}^{B_0(t_i-t)}+A_0}}\,,\label{rhoH}
\end{equation}
with $0<t_i$ the fixed time at the beginning of inflation and $C_0$ a fixed parameter ($[C_0]=[A_0]=[\kappa^2/H]$). When $(t_i-t)\ll1/B_0$, one gets the quasi de Sitter solution,
\begin{equation}
H\simeq H_0-\frac{3}{2\kappa^2}H_0^3 C_0 (t_i-t)\,,\label{HubbledS}
\end{equation}
where
\begin{equation}
H_0=\sqrt{\frac{\kappa^2}{3}}\sqrt{\frac{B_0}{C_0+A_0}}\quad B_0\,,C_0<0\,,\quad A_0<-C_0\,.
\end{equation}
In the above expressions, we see that $C_0\,,B_0$ must be negative in order to have
the slow decreasing of the (real) Hubble parameter during inflation.
 The inflation ends after the time
\begin{equation}
\Delta t\simeq-\frac{2\kappa^2}{3H_0^2 C_0}\,.
\end{equation}
We are assuming that $\Delta t<-1/B_0$ during inflation:
the Hubble parameter in (\ref{HubbledS}) has the typical values on inflation scale and 
to small value of $|B_0|$ corresponds small value of $|C_0+A_0|$.

Now we can calculate the $N$-folds number (\ref{N}) as
\begin{equation}
N\equiv\int^{t_f}_{t_i} H dt\simeq H_0\Delta t=-\frac{2\kappa^2}{3H_0 C_0}\simeq\frac{|\dot H(t_i)|}{H_0^2}\,.
\end{equation}
This expression holds true in the case of fluid models, where $N\sim 1/\epsilon$, as we have remarked in \S \ref{secomega}.
The duration of inflation must be at least of $\Delta t=76H_0^{-1}$, namely $-C_0/\kappa^2<-(114H_0)^{-1}$ . 
The slow roll parameters (\ref{epsilon})--(\ref{eta}) result to be
\begin{equation}
\epsilon\simeq-\frac{3H_0C_0}{2\kappa^2}\text{e}^{B_0(t_i-t)}\,,\quad
\eta\simeq\frac{B_0}{2H_0}-\frac{9H_0C_0}{4\kappa^2}\text{e}^{B_0(t_i-t)}\,,
\end{equation}
such that during inflation
\begin{equation}
\epsilon\simeq\frac{1}{N}\ll 1\,,\quad |\eta|\simeq
-\frac{3}{2N}\ll 1\,,
\end{equation}
where we have considered $B_0\ll H_0$ in the expression for $\eta$. Therefore, the slow roll conditions are well satisfied. 
The power spectrum (\ref{spectrum}) reads
\begin{equation}
\Delta_\mathcal{R}^2=\frac{\kappa^2 H_0^2 N}{8\pi^2}\,,
\end{equation} 
and the spectral indexes (\ref{index}) are given by
\begin{equation}
n_s=1-\frac{9}{N}
\,,\quad r=\frac{16}{N}\,.
\end{equation}
For $N=76$, this indexes read $n_s=0.882$ and $r=0.211$ and the results of the last  BICEP2 experiment can be realized from the model. On the other side, for larger values of the $N$-folds number we obtain $n_s$ and $r$ closer to one and zero, respectively, in the range of the Planck satellite data. 

At the end of inflation, the fluid model enters in a decelerated phase (when $\epsilon>1$):
therefore, solution (\ref{rhoH}) reads in the limit $(t_i-t)\gg 1/B_0$, $B_0<0$,
\begin{equation}
\rho\simeq\frac{B_0}{C_0}\text{e}^{B_0 t}\,,\quad 
H\simeq\sqrt{\frac{\kappa^2}{3}}\sqrt{\frac{B_0}{C_0}}\text{e}^{\frac{B_0 t}{2}}\,,
\end{equation}
which implies $\rho\sim\log[a(t)/a_0]$, where $a(t)$ has a minimum at $a(t)\simeq a_0$ when $t\rightarrow\infty$. The fluid energy density disappears in expanding universe, but a process of creation of particle is necessary to recover the Friedmann universe. In \S~\ref{secomega} we have discussed the problem, showing how with a suitable viscous term appearing at some small Hubble parameter scale, oscillations of the fluid can generate the oscillations of the curvature at the origin of the reheating. Similar process can be extended to any fluid. For example, for the fluid under consideration in this paragraph, it is enough to add to the viscosity (\ref{ff}) some counterterms of the Hubble parameter (remember that $\rho=3H^2/\kappa^2$), to cancel the pressure (\ref{ppp}) and reduce the analysis to the one of \S~\ref{secomega}.

\section{Inhomogeneous fluids for early- and late-time acceleration}

Many attempts have been done to unify inflation and the dark energy 
dominated epoch of current cosmic acceleration (see for example the first attempt in $F(R)$-gravity in Ref.~\cite{Odunification}). 
Here, we would like to conclude our work at a qualitative level, showing how it is possible to obtain an unified description with an inhomogeneous fluid-representation.

A toy model can be recovered by starting from (\ref{ooooo}), which is in fact a generalization of (\ref{o1}). Let us suppose to have an inhomogeneous fluid whose EoS-parameter is given by
\begin{equation}
\omega(\rho)=\left[-(\omega_\text{eff}+1)\frac{\rho}{\rho+\rho_*}+\omega_\text{eff}\right]\left[1-\left(\frac{\omega_\text{eff}-1}{\omega_\text{eff}}\right)\text{e}^{-\rho/\rho_{**}}\right]\,,\quad
\omega_\text{eff}\neq -1\,,
\end{equation}
where $\rho_*(=3H_*^2/\kappa^2)$ is an energy density on the inflation scale and $\rho_{**}(=3H_*^2/\kappa^2)\ll\rho_*$. In this way one has
\begin{equation}
\omega(\rho_*\ll\rho)\simeq -1\,,\quad
\omega(\rho_{**}\ll\rho\ll\rho_*)\simeq\omega_\text{eff}\,,\quad
\omega(\rho\ll\rho_{**})\simeq-1\,.
\end{equation}
Therefore, inflation is reproduced in the limit $\rho_*\ll\rho$, as it has been shown in \S {\ref{secomega}}. At the end of inflation, the Equation of State of the fluid changes and $\omega\simeq\omega_\text{eff}$ as long as $\rho_{**}\ll\rho\ll\rho_*$. The choice $\omega_\text{eff}=1/3$ can justify the presence of radiation/ultrarelativistic matter in the universe after inflation: otherwise, the reheating is necessary and a further analysis is required. Finally, when $\rho\ll\rho_{**}$, the fluid Equation of State tourns out to be the one of the dark energy and the current cosmic acceleration can be found.

\section{Conclusions}

In the present work, we have investigated the inflation in the inhomogeneous viscous fluid representation. Such a general choice of the Equation of State of the fluid permits to realize many cosmological scenarios. In our analysis, we looked for three toy models realizing inflation with different mechanisms, namely with an inhomogeneous EoS parameter whose behaviour changes at the end of inflation turning out the accelerated solution in a decelerated one  (\S~\ref{secomega}), a viscous fluid whose viscosity changes the fluid Equation of State at the end of inflation bringing the universe in a decelerated phase (\S~\ref{secviscous}), and a fluid whose Equation of State permits to find an exact solution which passes from an accelerated to a decelerated phase (\S~\ref{seclast}). Inflation can be realized from this models in a viable way: solution is always near to the de Sitter one, and the slow roll paramters are small. We found some differences with respect to the scalar theories for inflation, where the slow roll paramters result to be much smaller: typically in these theories the $\epsilon$ slow roll parameter behaves like $1/N^2$, while for fluids it goes like $1/N$, $N$ being the e-folds number of inflation. As a consequence, the fluid models permit to have de Sitter solutions more perturbated and the spectral index and the tensor to scalar ratio further to one and zero, respectively. This fact may be interesting expecially after the very new BICEP2 experiment results. 

In our models we payed attention also on the end of inflation. Thanks to the viscosity we can reconstruct the mechanism at the origin of the reheating, namely we may generate the oscillations of the Ricci scalar during a fluid dominated universe: in this way, the particle production can follow the inflation with or without the coupling of the curvature with the matter fields. An other possibility is given by the conversion of the fluid energy density in radiation one, such that the Friedmann universe contents come from the fluid itself. 

In the last section of this work, we presented some considerations about the possibility to unify the early-time acceleration with the last-time acceleration bringing this effects togheter in an unique `dark' fluid model.

Other works on inhomogeneous viscous fluids and the dark energy issue have been carried out in Ref.~\cite{Ciappi}, in Refs.~\cite{mio1}--\cite{dodici}, in Ref.~\cite{LittleRip} for viscous fluids in Little Rip cosmology, in Ref.~\cite{Carro} for other fluid interactions and in Ref.~\cite{pp} for fluid perturbations in FRW universe.



\begin{thebibliography}{}


\bibitem{WMAP}
Komatsu, E.; Dunkley, J.; Nolta, M.R.; Bennett, C.L.; Gold, B.; Hinshaw, G.; Jarosik, N.; Larson,~D.; Limon, M.; Page, L.; \emph{et al}. 
\emph{Astrophys. J. Suppl.} (2009), \emph{180}, {330--376}.



\bibitem{Linde}
A.~D.~Linde,
   Lect.\ Notes Phys.\  {\bf 738}, 1 (2008)
   [arXiv:0705.0164 [hep-th]].

\bibitem{revinflazione}
D.S.Gorbunov and V.A.Rubakov,
\textit{Introduction to the Theory of the Early Universe: Hot Big Bang Theory} 
(2011).


\bibitem{reviewmod}
 S.~Nojiri and S.~D.~Odintsov,
   Phys.\ Rept.\  {\bf 505}, 59 (2011)
   [arXiv:1011.0544 [gr-qc]].

\bibitem{reviewmod2}
 S.~Nojiri and S.~D.~Odintsov,
  eConf C {\bf 0602061}, 06 (2006)
  [Int.\ J.\ Geom.\ Meth.\ Mod.\ Phys.\  {\bf 4}, 115 (2007)]
  [hep-th/0601213].

\bibitem{myrev}
R.~Myrzakulov, L.~Sebastiani and S.~Zerbini,
  Int.\ J.\ Mod.\ Phys.\ D {\bf 22}, 1330017 (2013)
  [arXiv:1302.4646 [gr-qc]].


\bibitem{Guth}
A. H. Guth, Phys. Rev. D {\bf 23}, 347 (1981).

\bibitem{Sato}
K. Sato, Mon. Not. R. Astron. Soc. {\bf 195}, 467 (1981);  Phys. Lett. {\bf 
99B}, 66 (1981).

\bibitem{chaotic}
A. Linde, Phys. Lett. {\bf 129B}, 177 (1983).


\bibitem{buca1}
A. Linde, Phys. Lett. {\bf 108B}, 389 (1982).

\bibitem{buca2}
A. Albrecht and P. Steinhardt, Phys. Rev. Lett. {\bf 48}, 1220 (1982).

\bibitem{buca3}
K. Freese, J. A. Frieman, A. V. Orinto, Phys. Rev. Lett. {\bf 65}, 3233 (1990).

\bibitem{buca4}
F. C. Adams, J. R. Bond, K. Freese, J. A. Frieman and A. V. Orinto, Phys. Rev. 
D {\bf 47}, 426 (1993).

\bibitem{ibrida1}
A. D. Linde, Phys. Rev. D {\bf 49}, 748 (1994).

\bibitem{ibrida2}
E. J. Copeland, A. R. Liddle, D.H. Lyth, E. D. Stewart and D. Wands, Phys. Rev. 
D {\bf 49}, 6410 (1994).


\bibitem{Staro}
   A.~A.~Starobinsky,
   Phys.\ Lett.\ B {\bf 91}, 99 (1980).


\bibitem{Planckdata}
P.~A.~R.~Ade {\it et al.} [Planck Collaboration],
arXiv:1303.5076 [astro-ph.CO].

\bibitem{BICEP2} 
  P.~A.~R.~Ade {\it et al.}  [BICEP2 Collaboration],
  arXiv:1403.3985 [astro-ph.CO].



\bibitem{Staromio}
L.~Sebastiani, G.~Cognola, R.~Myrzakulov, S.~D.~Odintsov and S.~Zerbini,
  Phys.\ Rev.\ D {\bf 89}, 023518 (2014)
  [arXiv:1311.0744 [gr-qc]].

\bibitem{Staromio2}
K.~Bamba, R.~Myrzakulov, S.~D.~Odintsov and L.~Sebastiani,
  Phys.\ Rev.\ D {\bf 90}, 043505 (2014)
  [arXiv:1403.6649 [hep-th]].




\bibitem{fluidOd} 
 Capozziello, S.; Cardone, V.F.; Elizalde, E.; Nojiri, S.; Odintsov, S.D. 
\emph{Phys. Rev. D} \textbf{2006}, \emph{73}, 043512:1--043512:16.

\bibitem{fluidOd2}
 S.~Nojiri and S.~D.~Odintsov,
  Phys.\ Rev.\ D {\bf 72} (2005) 023003
  [hep-th/0505215].

\bibitem{fluidOd3}
 S.~Nojiri and S.~D.~Odintsov,
  Phys.\ Lett.\ B {\bf 639} (2006) 144
  [hep-th/0606025].

\bibitem{fluidOd4}
 K.~Bamba, S.~Capozziello, S.~'i.~Nojiri and S.~D.~Odintsov,
  Astrophys.\ Space Sci.\  {\bf 342}, 155 (2012)
  [arXiv:1205.3421 [gr-qc]].




\bibitem{Alessia} 
I.~H.~Brevik and O.~Gorbunova,
Gen.Rel.Grav. {\bf 37} 2039-2045 (2005). 

\bibitem{Brevik} 
I.~H.~Brevik and O.~Gorbunova,
  Eur.\ Phys.\ J.\ C {\bf 56}, 425 (2008)
  [arXiv:0806.1399 [gr-qc]].

\bibitem{Alessia(2)} 
I.~H.~Brevik, O.~Gorbunova and Y.A. Shaido, 
Int.\ J.\ Mod.\ Phys.\  D {\bf 14} 1899 (2005).


\bibitem{Caprev}
  S.~Capozziello and M.~De Laurentis,
  Phys.\ Rept.\  {\bf 509}, 167 (2011)
  [arXiv:1108.6266 [gr-qc]].




\bibitem{deFelice}
A.~De Felice and S.~Tsujikawa,
  Living Rev.\ Rel.\  {\bf 13}, 3 (2010)
  [arXiv:1002.4928 [gr-qc]].



\bibitem{davood}
M.~R.~Setare, D.~Momeni, V.~Kamali and R.~Myrzakulov,
  arXiv:1409.3200 [physics.gen-ph].


\bibitem{Odunification}
S.Nojiri and S.~D.~Odintsov,
Phys.\ Rev.\  {\bf D68}, 123512 (2003)
[hep-th/0307288].


\bibitem{Ciappi}
A.~Y.~.Kamenshchik, U.~Moschella and V.~Pasquier,
  Phys.\ Lett.\ B {\bf 511}, 265 (2001)
  [gr-qc/0103004].

\bibitem{mio1}
 O.~Gorbunova and L.~Sebastiani,
  Gen.\ Rel.\ Grav.\  {\bf 42}, 2873 (2010)
  [arXiv:1004.1505 [gr-qc]].

\bibitem{mio2}
 L.~Sebastiani,
  Eur.\ Phys.\ J.\ C {\bf 69}, 547 (2010)
  [arXiv:1006.1610 [gr-qc]].

\bibitem{mio3}
R.~Myrzakulov, L.~Sebastiani and S.~Zerbini,
  Galaxies {\bf 1}, no. 2, 83 (2013)
  [arXiv:1307.4854 [gr-qc]].

\bibitem{mio4}
S.~Myrzakul, R.~Myrzakulov and L.~Sebastiani,
  Astrophys.\ Space Sci.\  {\bf 350}, 845 (2014)
  [arXiv:1311.6939 [gr-qc]].

\bibitem{mio5}
R.~Myrzakulov and L.~Sebastiani,
  Astrophys.\ Space Sci.\  {\bf 352}, 281 (2014)
  [arXiv:1403.0681 [gr-qc]].

\bibitem{mio6}
 S.~Myrzakul, R.~Myrzakulov and L.~Sebastiani,
  Astrophys.\ Space Sci.\  {\bf 353}, 667 (2014)
  [arXiv:1406.1576 [gr-qc]].


\bibitem{uno} 
Cardone, V.F.; Tortora, C.; Troisi, A.; Capozziello, S. 
\emph{Phys. Rev. D} (2006), \emph{73}, 043508:1--043508:15.

\bibitem{due}
 S.~Nojiri and S.~D.~Odintsov,
  Phys.\ Lett.\ B {\bf 649} (2007) 440
  [hep-th/0702031 [HEP-TH]].

\bibitem{tre} 
 I.~Brevik, S.~Nojiri, S.~D.~Odintsov and D.~Saez-Gomez,
  Eur.\ Phys.\ J.\ C {\bf 69} (2010) 563
  [arXiv:1002.1942 [hep-th]].
  
\bibitem{quattro} I. Brevik and S.D. Odintsov,
\emph{Phys. Rev. D} (2002), {\bf 65}, 067302:1--067302:4.

\bibitem{cinque} D. Youm,
\emph{Phys. Lett. B} (2002), \emph{531}, {276--280}.

\bibitem{sei}
I.~Brevik, V.~V.~Obukhov and A.~V.~Timoshkin,
  arXiv:1410.2750 [gr-qc].

\bibitem{sette}
 N.~Majd and D.~Momeni,
  Int.\ J.\ Mod.\ Phys.\ E {\bf 20}, 113 (2011)
  [arXiv:0903.2020 [gr-qc]].

\bibitem{otto}
M.~Jamil, K.~Yesmakhanova, D.~Momeni and R.~Myrzakulov,
  Central Eur.\ J.\ Phys.\  {\bf 10}, 1065 (2012)
  [arXiv:1207.2735 [gr-qc]].

\bibitem{nove}
D.~Momeni, N.~Majd and R.~Myrzakulov,
  Europhys.\ Lett.\  {\bf 97}, 61001 (2012)
  [arXiv:1204.1246 [hep-th]].

\bibitem{dieci}
M.~R.~Setare and D.~Momeni,
  Int.\ J.\ Theor.\ Phys.\  {\bf 50}, 106 (2011)
  [arXiv:1001.3767 [physics.gen-ph]].

\bibitem{undici}
D.~Momeni and M.~R.~Setare,
  Mod.\ Phys.\ Lett.\ A {\bf 26}, 2889 (2011)
  [arXiv:1106.0431 [physics.gen-ph]].

\bibitem{dodici}
M.~Jamil, D.~Momeni and R.~Myrzakulov,
  Gen.\ Rel.\ Grav.\  {\bf 45}, 263 (2013)
  [arXiv:1211.3740 [physics.gen-ph]].

\bibitem{LittleRip}
I.~Brevik, E.~Elizalde, S.~Nojiri and S.~D.~Odintsov,
  Phys.\ Rev.\ D {\bf 84} (2011) 103508
  [arXiv:1107.4642 [hep-th]].

\bibitem{Carro}
S.~M.~Carroll, M.~Hoffman and M.~Trodden,
  Phys.\ Rev.\ D {\bf 68}, 023509 (2003)
  [astro-ph/0301273].

\bibitem{pp}
 A.~V.~Astashenok and S.~D.~Odintsov,
  Phys.\ Lett.\ B {\bf 718}, 1194 (2013)
  [arXiv:1211.1888 [gr-qc]].



\end{thebibliography}
\end{document}